\documentclass{article}
\usepackage{spconf,amsmath,graphicx}
\usepackage{amsfonts, pifont}
\usepackage{multirow, tabu, booktabs}
\usepackage{xcolor, colortbl}
\usepackage{url}
\usepackage{bm}
\usepackage{hyperref}
\newcommand{\cmark}{\ding{51}}%
\newcommand{\xmark}{\ding{55}}%

\title{On the Importance of Neural Wiener Filter for Resource Efficient Multichannel Speech Enhancement}
\name{
 Tsun-An Hsieh$^\dag$,
 Jacob Donley$^\ddagger$,
 Daniel Wong$^\ddagger$,
 Buye Xu$^\ddagger$,
 Ashutosh Pandey$^\ddagger$
 }
\address{
 $^\dag$Department of Intelligent Systems Engineering, Indiana University Bloomington, IN, USA\\
 $^\ddagger$Reality Labs Research, Meta, WA, USA
}
\begin{document}
\ninept
\maketitle
\begin{abstract}
We introduce a time-domain framework for efficient multichannel speech enhancement, emphasizing low latency and computational efficiency. This framework incorporates two compact deep neural networks (DNNs) surrounding a multichannel neural Wiener filter (NWF).
The first DNN enhances the speech signal to estimate NWF coefficients, while the second DNN refines the output from the NWF. The NWF, while conceptually similar to the traditional frequency-domain Wiener filter, undergoes a training process optimized for low-latency speech enhancement, involving fine-tuning of both analysis and synthesis transforms.
Our research results illustrate that the NWF output, having minimal nonlinear distortions, attains performance levels akin to those of the first DNN, deviating from conventional Wiener filter paradigms. Training all components jointly outperforms sequential training, despite its simplicity. Consequently, this framework achieves superior performance with fewer parameters and reduced computational demands, making it a compelling solution for resource-efficient multichannel speech enhancement.
\end{abstract}
\section{Introduction}
\label{sec:intro}
Sequential neural beamforming has gained prominence as a potent technique for multichannel speech enhancement (SE), offering substantial improvements in performance \cite{qian2018deep, luo2019fasnet, luo2022time, wang2022stft, lu2022towards, kuang2023three, lee2023improved, wang2023tf}. It generally contributes to an increased robustness of downstream applications, such as automatic speech recognition (ASR) \cite{wang2018spatial, ochiai2017unified, zhang2022end, wang2023tf} and speaker verification \cite{movsner2022multi, dowerah2022compensating}.

\begin{figure}[h!]
  \centering
  \centerline{\includegraphics[width=0.76\linewidth]{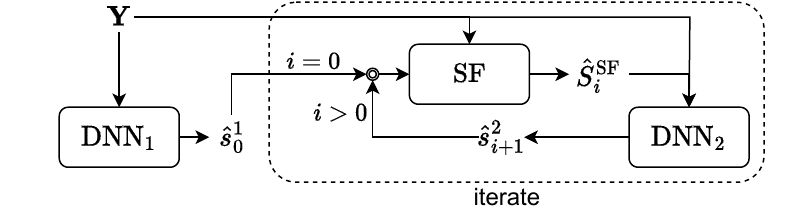}}
%  \vspace{2.0cm}
  % \centerline{(a) Result 1}\medskip
%
\caption{A general workflow of sequential neural beamforming.}
\label{fig:seq_bf}
\end{figure}

Sequential neural beamforming is characterized by a series of iterative steps. Initially, an enhancement network processes the multichannel audio input, extracting a less distorted speech signal. This processed signal serves as the target for estimating the spatial filter in the subsequent step, with the aim of further reducing nonlinear distortions. Subsequently, both the noisy mixture and the spatial filter estimate are passed to a second-stage enhancement network, culminating in the generation of the final enhanced speech. Optionally, the sequence of spatial filter estimation followed by enhancement network processing can be repeated multiple times to gradually improve the performance. Fig. \ref{fig:seq_bf} provides a visual overview of sequential neural beamforming.

Typically, sequential neural beamforming relies on underlying DNNs operating in the frequency-domain to enhance the short-time Fourier transform (STFT) of noisy speech \cite{wang2020multi, wang2022stft, wang2023tf, wang2023fneural}. In addition, these systems often incorporate traditional spatial filters such as the multichannel Wiener filter (MCWF), minimum-variance distortionless response (MVDR) \cite{doclo2010acoustic}, and generalized eigenvalue (GEV) \cite{warsitz2007blind} beamformers. These traditional filters are valuable because they can capture spectral patterns critical for isolating overlapping sources or dealing with reverberation. However, it is important to note that they have limitations, including numerical instability due to matrix inversion and the fact that filter coefficients are derived independently by solving a minimum mean squared error (MMSE) problem. As a result, jointly optimizing frequency-domain beamformers with deep neural networks (DNNs) can be challenging.  To address the training instability associated with traditional spatial filters, a widely accepted strategy is to employ diagonal loading ~\cite{mestre2003diagonal}. Furthermore, a recent study introduced the use of recurrent neural networks to directly estimate the matrix inverse, offering an alternative approach to mitigate training challenges ~\cite{zhang2021adl}.

In contrast, the advent of end-to-end systems has prompted some researchers to consider time-domain spatial filters as trainable components that can be jointly optimized with neural networks ~\cite{qian2018deep, luo2019fasnet, luo2022time}. For instance, the filter-and-sum network (FaSNet) ~\cite{luo2019fasnet} employs a set of learnable filters that convolve with multichannel inputs, and their outputs are summed to produce the beamformed output. However, due to its end-to-end design, FaSNet may not guarantee certain desirable properties of traditional filters, such as a distortionless response or temporal consistency in filtering. A recent study by Luo et al. ~\cite{luo2022time} addresses this limitation by introducing a time-domain generalized Wiener filter that derives filter coefficients using MMSE solution over a trainable latent representation. 

While sequential neural beamforming has demonstrated effectiveness, the majority of research has been conducted in a non-causal setting, often neglecting resource constraints like computational demands, algorithmic latency, and model size \cite{luo2022time, wang2023tf, wang2020multi}. While a few studies in the frequency domain, such as \cite{wang2022stft, lu2022towards, wang2023fneural}, have started considering these aspects, the exploration of end-to-end optimization in the time domain remains relatively unexplored. This is especially noteworthy, given the substantial potential of end-to-end optimization for resource-constrained speech enhancement \cite{zhang2020end, pandey23b_interspeech, patel2023ux}.

We propose a novel time-domain sequential neural beamforming framework for resource efficient multichannel speech enhancement. This framework operates with an impressively low algorithmic latency of just 2 milliseconds, demonstrating remarkable efficiency in terms of both model parameters and computational demands. Central to our system are two lightweight, low-latency recurrent neural networks (LLRNNs) \cite{pandey23b_interspeech} employed as key DNN components, as depicted in Fig. \ref{fig:seq_bf}. These LLRNNs collaborate to simultaneously suppress noise and reverberation. Additionally, we introduce a novel neural Wiener filter (NWF) to function as the spatial filter, identified as the SF block in Fig. \ref{fig:seq_bf}.

Given a multichannel noisy speech, the first-stage DNN ($\mathrm{DNN}_1$) produces a single-channel enhanced speech that serves as the target signal for estimating NWF coefficients. The NWF conceptually resembles a traditional frequency-domain MCWF, with the key distinction being that its analysis and synthesis transforms are trained alongside other components. This adaptability allows these transforms to be fine-tuned for the specific requirements of low-latency speech enhancement, a task that may pose challenges for conventional filters. Subsequently, the multichannel noisy input is processed by the NWF, yielding a less distorted single-channel speech. The NWF output is then combined with the original input and fed to the second DNN ($\mathrm{DNN}_2$) to obtain the final enhanced speech. Different blocks in the framework are setup in a way that stacking multiple blocks do not result in increased algorithmic latency.

To identify the optimal training approach, we conducted an extensive exploration of different training strategies, including various training orders for each module, combinations of pretrained weights, initialization methods, and loss configurations. We also provide extensive comparisons with competitive baseline models to showcase the promise of proposed framework for resource efficient speech enhancement. 

%Our research findings highlight that training all system components end-to-end, with random initialization of the neural Wiener filter (NWF) and applying a loss function at the final output, delivers the most promising results. 

%In our experimental setups, our optimized system showcases superior performance compared to the LLRNN baseline, which shares the same hidden size of 128. Specifically, our system outperforms the baseline by 6.62\% in terms of short-time objective intelligibility (STOI) and achieves a 3.14dB improvement in scale-invariant source-to-distortion ratio (SI-SDR) \cite{le2019sdr}.

%Moreover, when considering models with a similar scale of parameters, our proposed system continues to outshine the baseline, surpassing it by 3.52\% in STOI and achieving a 1.84 dB higher SI-SDR. Importantly, despite delivering competitive performance, the baseline model demands approximately 4.22 times the number of parameters and 2.23 times the floating-point operations per second (FLOPS) in comparison to our proposed method. This highlights the computational efficiency and low resource requirements of our approach.
% 
\begin{figure}[tb]
  \centering
  \centerline{\includegraphics[width=\linewidth]{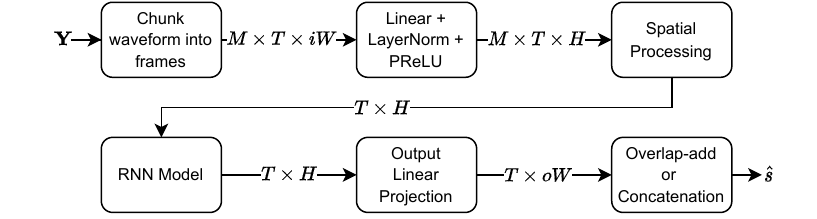}}
%  \vspace{2.0cm}
  % \centerline{(a) Result 1}\medskip
%
\caption{Flow chart of LLRNN.}
\label{fig:llrnn}
\end{figure}
\section{Proposed Method}
\label{sec:method}
\subsection{Problem Formulation}
\label{ssec:seq_bf}
A multichannel signal recorded using an array with $M$ microphones in a noisy and reverberant environment can be defined as $\mathbf{Y}=\{\bm{y}_m\}^M_{m=1}$, which includes $M$ observations at  $M$ microphones. The $m^{\rm th}$ observation $y_m\in \mathbb{R}^{1\times L}$ with $L$ samples can be decomposed as:
\begin{align}
    \bm{y}_m = \bm{s}_{m}^{dir} + \bm{s}_{m}^{rev} + \bm{z}_{m}^{dir} + \bm{z}_{m}^{rev}
\end{align}
where $\bm{s}_{m}^{dir}$, $\bm{s}_{m}^{rev}$, $\bm{z}_{m}^{dir}$, and $\bm{z}_{m}^{rev}$ respectively represent direct path speech, speech reverberation, direct path noise, and noise reverberation. The goal of multichannel speech enhancement system is get a close estimate of direct path speech $\bm{s}_{r}^{dir}$ at a given reference mic $r$. 

\subsection{Sequential Neural Beamforming}
A general framework of sequential neural beamforming is shown in Fig. ~\ref{fig:seq_bf}. Superscripts and subscripts respectively indicate the DNN and iteration index. Given a noisy mixture $\mathbf{Y}$, $\mathrm{DNN}_1$ estimates an intermediate enhanced speech $\hat{\bm{s}}^{1}_0$. Next, a spatial filter uses $\hat{\bm{s}}^{1}_0$ as the target to obtain the beamformed speech $\hat{\bm{s}}^{\rm SF}_0$, which is then concatenated with $\mathbf{Y}$ across channel and processed by $\mathrm{DNN}_2$ to obtain $\hat{\bm{s}}^{2}_1$. The output of the spatial filter contains minimal nonlinear distortions, and as a result, $\mathrm{DNN}_2$ is expected to produce $\hat{\bm{s}}^{2}_1$ with improved nonlinear distortions. Optionally, the stack of spatial filter and $\mathrm{DNN}_2$ can be repeated for $i\geq 1$ to obtain $\hat{\bm{s}}^{\rm SF}_i$ and $\hat{\bm{s}}^{2}_{i+1}$.

\subsection{Lightweight Low-latency RNN}
\label{ssec:se_model}
The proposed framework incorporates two deep neural networks (DNNs), both of which are built upon a lightweight, low-latency Recurrent Neural Network (LLRNN) architecture introduced in \cite{pandey23b_interspeech}. Illustrated in Fig. \ref{fig:llrnn}, the LLRNN serves as a time-domain, multichannel speech enhancement model for processing noisy signal $\mathbf{Y} \in \mathbb{R}^{M \times L}$. The processing begins by converting $\mathbf{Y}$ into a sequence of overlapping frames, employing a frame size of $iW$ and a frame shift of $J$, resulting in $\bar{\bm{Y}} \in \mathbb{R}^{M \times T \times iW}$, where $T$ signifies the number of frames. Prior to frame conversion, the signal is padded with $iW-J$ zeros at the beginning. Subsequently, a linear layer is applied, followed by layer normalization \cite{ba2016layer} and parametric rectified linear units (PReLU) \cite{he2015delving}, to project all frames into a latent representation of size $H$. 

Following this, a spatial processing block is employed to reduce the channel dimension, mapping input of size $M \times T \times H$ to $T \times H$. The reduced feature tensors then traverse through $B$ consecutive recurrent blocks, each comprising layer normalization followed by a Long Short-Term Memory (LSTM) \cite{hochreiter1997long} with a hidden size of $H$. Subsequently, all frames are linearly projected to the output frame size of $oW$ using a linear layer, producing the sequence of enhanced frames. 

The choice of the output frame size, whether it is $oW$ or $J$ (frame shift), depends on the specific stage of the LLRNN. If the LLRNN output serves as the final output for evaluation, the output frame size is set to $oW$, and overlap-add (OLA) is applied to obtain the enhanced signal. In this scenario, the LLRNN exhibits an algorithmic latency of $oW$, as $oW$ samples in the output frame correspond to the rightmost $oW$ samples in the input frame \cite{pandey23b_interspeech}. Alternatively, if the LLRNN output is fed into a subsequent model, the output frame size is set to $J$, and the enhanced signal is obtained by simply concatenating the output frames. This particular setup operates with an algorithmic latency of $J$. Importantly, in this setup, stacking multiple LLRNNs does not result in an increase in overall algorithmic latency due to the absence of OLA at intermediate stages.

%We adopt a lightweight low-latency RNN (LLRNN) proposed in \cite{pandey23b_interspeech} as the first and second DNN in the proposed framework. As shown in Fig. \ref{fig:llrnn}, LLRNN is a time-domain multichannel speech enhancement model taking $\mathbf{Y} \in \mathbb{R}^{M \times L}$ as input. It first converts $\mathbf{Y}$ into a sequence of overlapping frames $\bar{Y} \in \mathbb{R}^{M \times T \times iW}$, where $T$ and $iW$ are the number of frames and the input frame size, respectively. Next, it utilizes a linear layer followed by layer normalization \cite{ba2016layer} and parametric rectified linear unit (PReLU) \cite{he2015delving} to project all the frames to a latent representation of size $H$. After this, a spatial processing block is used to collapse the channel dimension, by mapping input of size $M \times T \times H$ to $T \times H$. The reduced feature tensors are then processed by $B$ consecutive recurrent blocks that comprises a layer normalization followed by a long short-term memory (LSTM) \cite{hochreiter1997long} of hidden size $H$, and then all of the frames are then projected to the output frame size of $oW$ using a linear layer to obtain the sequence of enhanced frames, which is consequently converted back into a single channel waveform $\hat{s} \in \mathbb{R}^{1 \times L}$ using overlap-add (OLA) or concatenation, depending on the stage LLRNN belonging to. Details of waveform reconstruction methods will be discussed in \ref{ssec:model_config}.

% 
\begin{figure}[tb]
  \centering
  \centerline{\includegraphics[width=\columnwidth]{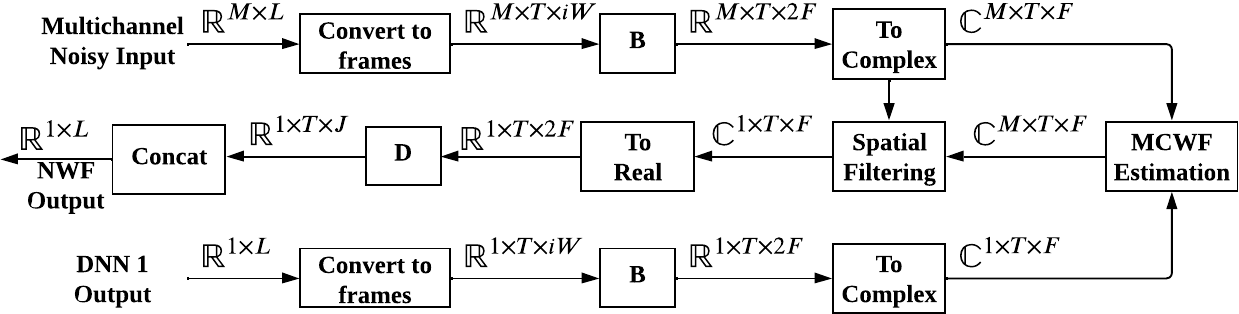}}
%  \vspace{2.0cm}
  % \centerline{(a) Result 1}\medskip
%
\caption{Flow chart of the proposed NWF.}
\label{fig:td_gwf}
\end{figure}
\subsection{Neural Wiener Filter}
\label{ssec:beamformer}
Fig. \ref{fig:td_gwf} shows the pipeline of our proposed NWF. The multichannel noisy input and the output from the first DNN are first converted to frames using a frame size of $iW$ and frame shift of $J$. Both signals are padded with $iW - J$ zeros in the beginning before converting to frames. After this, they are projected to size $2F$ using analysis matrix $\rm \bold{B}$ and then converted to complex valued tensors by using the first half as the real part and the second half as the imaginary part. Next, MCWF filter coefficients are estimated using the representation of the DNN$_{1}$ output as the target. A detailed description on how to estimate MCWF coefficients can be found in \cite{wang2022stft}.

Next, spatial filtering is applied over the multichannel noisy transform to obtain a single-channel enhanced transform. The enhanced transform is converted to real-valued tensor by stacking real and imaginary parts side by side. Then, the synthesis matrix $\mathbf{D}$ is multiplied with the real-valued enhanced transform to get enhanced frames of size $J$. Finally, all the enhanced frames are concatenated to obtain the enhanced signal. The algorithmic latency of this setup is $J$. If the NWF output is used as the final signal for evaluation, then the output frame size is set to $oW$ and OLA is applied to obtain enhanced waveform with an algorithmic latency of $oW$. 

It is important to note that the setup described here differs from the one in \cite{luo2022time}, where a real-valued Wiener filter was utilized. We observed that the real-valued formulation resulted in subpar performance compared to the complex-valued one. Similar to \cite{wang2022stft} and \cite{higuchi2018frame}, the proposed NWF employs a frame-online approach utilizing Woodbury formula for matrix inversion during evaluation.

\section{Experimental Setup}
\label{sec:exp}

\subsection{Noisy Reverberant Data Simulation}
\label{ssec:data}
To generate pairs of clean and noisy signals for training, we make use of the Interspeech 2020 DNS Challenge corpus \cite{reddy2020interspeech}. The speakers in the training set are randomly divided into sets for training, testing, and validation, with a split ratio of 85\%, 5\%, and 10\%, respectively. Similarly, the noises are categorized into distinct sets for training, testing, and validation. All the utterances are resampled to $16$ kHz before being used for data generation. 

Multichannel data is generated using an eight-microphone circular array with a radius of 10 cm. The data generation follows an algorithm outlined in previous studies \cite{pandey2022multichannel, pandey2022tparn, pandey23b_interspeech}. We generate 80K training, 1.6K validation, and 3.2K test utterances.

We use Pyroomacoustics with an image method of order 6 to generate room impulse responses (RIRs). The dimensions of the rooms, including length, width, and height, are sampled uniformly from the ranges $[3, 10]$ meters (m), $[3, 10]$ m, and $[2, 5]$ m, respectively. The absorption coefficients are sampled from $[0.1, 0.4]$. The number of noise sources is sampled from $[1, 10]$. The signal-to-noise ratio (SNR) is uniformly sampled from $-10$ to $10$ dB, representing the ratio between the direct-path speech and interferences, excluding speech reverberation. The direct-path speech at the first microphone is used as the training target. 

% % 
% \begin{table}[t]
% \centering
% \caption{Model configurations and specifics. $iW$, $oW$, and the $lty$ denote the input window size, output window size, and the latency, respectively, and are measured in milliseconds.}
% \label{tab:model_config}
% \resizebox{\columnwidth}{!}{%
% \begin{tabular}{l|c c c c}
% \toprule
% Setup & $H$ & $iW$ & $oW$ & $lty$\\
% \midrule
% \multirow{2}{*}{Baseline} 
% & 128 & 16 & 2 & 2 \\
% & 200 & 16 & 2 & 2 \\
% % & 256 & 16 & 2 & 2 &  &  \\
% \midrule
% % 
% ${\rm DNN_1}$ & 128 & 15 & 1 & 1 \\
% %
% \midrule
% % 
% DNN$_1$ + DNN$_2$          & 128/128 & 15/16 & 1/2 & 2 \\
% DNN$_1$ + MCWF + DNN$_2$   & 128/128 & 15/16 & 1/2 & 2 \\
% DNN$_1$ + TD-GWF + DNN$_2$ & 128/128 & 15/16 & 1/2 & 2 \\
% \bottomrule
% \end{tabular}
% }
% \end{table}
% % 
\subsection{Model Configuration and Training}
\label{ssec:model_config}
In this work, we consider six setups, including: 
(i) single-stage baselines, 
(ii)  a DNN ($\mathrm{DNN}_1$) followed by an NWF, 
(iii) $\mathrm{DNN}_1$ followed by another DNN ($\mathrm{DNN}_2$) without an NWF
(iv) train $\mathrm{DNN}_1$ and then jointly train a stack of $\mathrm{DNN}_1$, NWF and $\mathrm{DNN}_2$, 
(v) train a stack of $\mathrm{DNN}_1$ and NWF together, and then jointly train a stack of $\mathrm{DNN}_1$, NWF and $\mathrm{DNN}_2$
(vi) train a stack of $\mathrm{DNN}_1$, NWF and $\mathrm{DNN}_2$ from scratch. These categories correspond to setups 1 to 6 in Table \ref{tab:result}.

In particular, $\mathrm{DNN}_1$ is an LLRNN model with a latency of $1$ ms, employing parameters $H=128$, $iW=256$, and $oW=J=16$. In (ii), we merge $\mathrm{DNN}_1$ with NWF and utilize the NWF output, where a $2$ ms latency is achieved by using $oW=32$ with OLA in the synthesis transform. Case (iii) represents a simple dual-stage setup, where $\mathrm{DNN}_1$ can be pretrained (or trained from scratch) and frozen (or unfrozen) when training with $\mathrm{DNN}_2$. Similarly, (iv) through (vi) incorporate $\mathrm{DNN}_1$ and $\mathrm{DNN}_2$ around NWF. Here, $\mathrm{DNN}_1$ and NWF use $oW=J=16$ and concatenate output frames, while $\mathrm{DNN}_2$ utilizes $oW=32$ with OLA. The NWF maintains $F=129$ across all cases.

In the aforementioned setups, multiple enhanced speech are generated, including those from $\mathrm{DNN}_1$, NWF, and $\mathrm{DNN}_2$. We conduct experiments to asses the effectiveness of minimizing loss at different outputs by simply adding different losses together.

We train various baseline models including LLRNN \cite{pandey23b_interspeech}, MC-Conv-TasNet \cite{zhang2020end}, MC-CRN \cite{wang2022stft}, UXNet \cite{patel2023ux}, and FSB-LSTM \cite{wang2023fneural}. These models are tailored for single-stage processing, exhibiting latencies of either $2$ ms or $4$ ms. The LLRNN baseline achieves a latency latency of $2$ ms by using $oW=32$ ($2$ ms) with OLA.

\begin{table*}[t]
\centering
\caption{Model configurations and performance are summarized below. The symbols \cmark and \xmark  respectively denote the trainable and frozen modules. In the loss column, $\mathrm{DNN}_1$, NWF, and $\mathrm{DNN}_2$ denote the outputs from these modules, utilized in calculating the loss. The total loss is derived by summing the losses from these modules.}
\label{tab:result}
\resizebox{0.76\linewidth}{!}{%
\begin{tabular}{c| l |c c c|c c c c|c c}
\toprule
\multicolumn{2}{c|}{Setup} & STOI (\%) & PESQ & SI-SDR (dB) & Update? & init$_{\rm NWF}$ & Loss & Latency (ms) & \#params (M) & GFLOPs \\
\midrule
0A & Unprocessed & 65.83 & 1.63 & -7.48 & \multicolumn{4}{c}{N/A} & \multicolumn{2}{c}{N/A} \\
\midrule
% \multicolumn{10}{c}{\cellcolor{gray!25}Baseline} \\
1A & LLRNN$_{H=128}$ & 80.8 & 2.27 & 2.9 & \multirow{12}{*}{\cmark} & \multirow{12}{*}{\scriptsize{N/A}} & \multirow{12}{*}{\scriptsize{$\mathrm{DNN}_1$}} & 2 & 0.44 & 1.34 \\
1B & LLRNN$_{H=200}$ & 83.9 & 2.43 & 4.2 &  &  & &  2 & 1.03 & 2.78 \\
1C & LLRNN$_{H=256}$ & 85.6 & 2.51 & 4.9 &  &  &  & 2 & 1.66 & 4.25 \\
1D & LLRNN$_{H=300}$ & 86.2 & 2.56 & 5.3 &  &  &  & 2 & 2.26 & 5.61 \\
1E & LLRNN$_{H=400}$ &  87.5 &  2.64 &  6.0 &  &  &  & 2 & 3.97 & 9.40 \\
1F & LLRNN$_{H=512}$ & 88.3 & 2.69 & 6.5 &  &  &  & 2 & 6.46 & 14.79 \\
1G & MC-Conv-TasNet \cite{zhang2020end} & 86.3 & 2.57 & 5.6 &  &  &  & 2 & 5.13 & 10.32 \\
1H & MC-CRN-2ms \cite{wang2022stft}     & 84.0 & 2.38 & 3.9 &  &  &  & 2 & 2.32 & 6.73 \\
1I & MC-CRN-4ms                         & 85.7 & 2.51 & 4.7 &  &  &  & 4 & 2.32 & 6.73 \\
1J & UXNet-128 \cite{patel2023ux}       & 77.3 & 2.10 & 1.1 &  &  &  & 2 & 0.21 & 0.67 \\
1K & UXNet-256                          & 80.9 & 2.25 & 2.9 &  &  &  & 2 & 0.81 & 2.12 \\
1L & FSB-LSTM \cite{wang2023fneural}    & 88.2 & 2.68 & 5.8 &  &  &  & 4 & 1.97 & 7.80 \\
\midrule
% \multicolumn{10}{c}{\cellcolor{gray!25}$\mathrm{DNN}_1$+SF} \\
2A & \multirow{6}{*}{\shortstack[c]{$\mathrm{DNN}_1$+NWF}} 
      & 75.7 & 2.00 & -1.3 & \cmark/\xmark & \scriptsize{DFT} & \scriptsize{$\mathrm{DNN}_1$}    & \multirow{6}{*}{1/2} & \multirow{6}{*}{0.51} & \multirow{6}{*}{2.82} \\
2B &  & 75.6 & 2.01 & -0.3 & \cmark/\xmark & \scriptsize{DFT} & \scriptsize{$\mathrm{DNN}_1$+NWF} &  &  &  \\
2C &  & 82.0 & 2.14 &  3.2 & \cmark/\cmark & \scriptsize{DFT} & \scriptsize{NWF} &  &  &  \\
2D &  & 80.1 & 2.22 &  1.5 & \cmark/\cmark & \scriptsize{DFT} & \scriptsize{$\mathrm{DNN}_1$+NWF} &  &  &  \\
2E &  & 82.0 & 2.18 &  3.4 & \cmark/\cmark & \scriptsize{Rand.}& \scriptsize{NWF} &  &  &  \\
2F &  & 80.4 & 2.20 &  1.9 & \cmark/\cmark & \scriptsize{Rand.} & \scriptsize{$\mathrm{DNN}_1$+NWF} &  &  &  \\
\midrule
3A & $\mathrm{DNN}_1$+$\mathrm{DNN}_2$   & 82.3 & 2.33 & 3.4 & \cmark/\cmark & \multirow{4}{*}{\scriptsize{N/A}} & \scriptsize{$\mathrm{DNN}_1$ + $\mathrm{DNN}_2$} & \multirow{4}{*}{1/2} & \multirow{4}{*}{0.87} & \multirow{4}{*}{2.74}  \\
3B & $\mathrm{DNN}_1$+$\mathrm{DNN}_2$   & 82.9 & 2.36 & 3.8 & \cmark/\cmark & & \scriptsize{$\mathrm{DNN}_2$} &  &  &  \\
3C & $\mathrm{DNN}_1^{\rm PT}$+$\mathrm{DNN}_2$ & 83.5 & 2.37 & 4.0 & \cmark/\cmark &  & \scriptsize{$\mathrm{DNN}_2$} &  &  &  \\
3D & $\mathrm{DNN}_1^{\rm PT}$+$\mathrm{DNN}_2$ & 81.3 & 2.26 & 2.9 & \xmark/\cmark &  & \scriptsize{$\mathrm{DNN}_2$} &  &  &  \\
\midrule
4A & \multirow{4}{*}{$\mathrm{DNN}_1^{\rm PT}$+NWF+$\mathrm{DNN}_2$} 
      & 82.6 & 2.36 & 3.0 & \xmark/\xmark/\cmark & \scriptsize{DFT}  & \multirow{4}{*}{\scriptsize{$\mathrm{DNN}_2$}} & \multirow{4}{*}{1/1/2} & \multirow{4}{*}{0.94} & \multirow{4}{*}{4.21} \\
4B &  & 82.3 & 2.31 & 3.0 & \cmark/\xmark/\cmark & \scriptsize{DFT}  &  &  &  &  \\
4C &  & 85.7 & 2.52 & 4.7 & \xmark/\cmark/\cmark & \scriptsize{Rand.} &  &  &  &  \\
4D &  & 86.4 & 2.52 & 5.6 & \cmark/\cmark/\cmark & \scriptsize{Rand.} &  &  &  &  \\
\midrule
5A & \multirow{4}{*}{($\mathrm{DNN}_1$+NWF)$^{\rm PT}$+$\mathrm{DNN}_2$} & 82.1 & 2.32 & 2.1 & \xmark/\xmark/\cmark & \scriptsize{DFT} & \multirow{4}{*}{\scriptsize{$\mathrm{DNN}_2$}} & \multirow{4}{*}{1/1/2} & \multirow{4}{*}{0.94} & \multirow{4}{*}{4.21} \\
5B &  & 84.5 & 2.44 & 3.8 & \xmark/\xmark/\cmark & \scriptsize{Rand.} &  &  &  &  \\
5C &  & 84.5 & 2.43 & 3.7 & \xmark/\cmark/\cmark & \scriptsize{Rand.} &  &  &  &  \\
5D &  & 86.0 & 2.52 & 5.3 & \cmark/\cmark/\cmark & \scriptsize{Rand.} &  &  &  &  \\
% 5E &  &  &  &  & \cmark/\xmark/\cmark & DFT &  &  &  &  \\
% 
\midrule
% 
% 6A & \multirow{4}{*}{$\mathrm{DNN}_1$+NWF+$\mathrm{DNN}_2$}              
%    &  &  &  & \cmark/\xmark/\cmark & DFT & stg$_2$ & \multirow{4}{*}{1/1/2} &  & \multirow{4}{*}{4.21} \\
6A & \multirow{3}{*}{$\mathrm{DNN}_1$+NWF+$\mathrm{DNN}_2$}              
   & 84.9 & 2.46 & 4.0 & \cmark/\cmark/\cmark & \scriptsize{Rand.} & \scriptsize{$\mathrm{DNN}_1$+NWF+$\mathrm{DNN}_2$} & \multirow{3}{*}{1/1/2} & \multirow{3}{*}{\textbf{0.94}} & \multirow{3}{*}{\textbf{4.21}} \\
% 6B &  & 84.86 & 2.46 & 4.04 & \cmark/\cmark/\cmark & Random & stg$_1$+sf+stg$_2$ &  &  &  \\
6B &  & 86.3 & 2.53 & 5.0 & \cmark/\cmark/\cmark & \scriptsize{Rand.} & \scriptsize{NWF+$\mathrm{DNN}_2$} &  &  &  \\
6C &  & \textbf{87.4} & \textbf{2.58} & \textbf{6.0} & \cmark/\cmark/\cmark & \scriptsize{Rand.} & \scriptsize{$\mathrm{DNN}_2$} &  &  &  \\
6D & DNN$_{1}^{200}$ + NWF + DNN$_{2}^{200}$ & \textbf{89.1} & \textbf{2.70} & \textbf{7.0} & \cmark/\cmark/\cmark & \scriptsize{Rand.} & \scriptsize{$\mathrm{DNN}_2$} & 1/1/2 & 2.12 &  7.14 \\
\bottomrule
\end{tabular}
}
\end{table*}
% 
% For dual-stage models, we use $\mathrm{DNN}_1$+NWF+$\mathrm{DNN}_2$ as an another baseline, and the last two rows in Table \ref{tab:model_config} are the studied neural beamformers. In a dual-stage model, $\mathrm{DNN}_2$ has the same structure with the baseline model, and uses OLA to convert frames back into waveform. This means that the performance of $\mathrm{DNN}_1$ + $\mathrm{DNN}_2$ should be better or at least equal to the baseline model since the latency of both systems are the same.

For model optimization, we train these setups for 200 epochs using Adam optimizer \cite{kingma2014adam} with a constant learning rate of $2\times 10^{-4}$ and {\it amsgrad} enabled.
Following \cite{pandey23b_interspeech}, we adopt the phase constrained magnitude (PCM) loss \cite{pandey2021dense} to train all the models. In order to prevent unstable gradients, we clip gradient's $L^2$-norm to 0.03. Training for all setups is carried out on Nvidia V100 GPUs with automatic mixed precision.
% 
% \begin{figure}[tb]
%   \centering
%   \centerline{\includegraphics[width=0.8\linewidth]{figs/large_ConcatVSOLA.pdf}}
% %  \vspace{2.0cm}
%   % \centerline{(a) Result 1}\medskip
% %
% \caption{Waveform reconstruction via (a) overlap-add and (b) concatenation.}
% \label{fig:concat_vs_ola}
%
% \end{figure}
% 
\subsection{Evaluation Metrics}
\label{ssec:metrics}
Models are evaluated using perceptual evaluation of quality (PESQ), scale-invariant source-to-distortion ratio (SI-SDR), and short-time objective intelligibility (STOI) scores. Higher scores indicate better performance. The amount of computation is reported in Giga floating point operations (GFLOPs) for processing one second of 8-channel speech.
% We evaluate our systems based on the quality and intelligibility. For quality evaluation, we adopt the perceptual evaluation of quality scores (PESQ) and the scale-invariant source-to-distortion ratio (SI-SDR). For intelligibility, short-time objective intelligibility (STOI) measure is utilized. For all metrics higher values mean better performance. .

\section{Results}
\label{sec:result}

\subsection{Baseline and Single-Stage Setups}
\label{ssec:single_stage}

Table \ref{tab:result} provides a comprehensive overview of performances and computational demands across all configurations. Scores from setup 0A pertain to the noisy reverberant mixture. An anticipated trend from 1A to 1F is evident, indicating improved performance with an increase in hidden size.

Setups 2A to 2F demonstrate the impact of combining $\mathrm{DNN}_1$ with NWF. In 2A and 2B,  NWF training is disabled, with signal transform initialized using DFT/iDFT coefficients, making them equivalent to traditional frequency-domain MCWF. Setups 2C to 2F employ NWF with trainable signal transforms. Notably, 2C and 2D initialize signal transforms with DFT coefficients, while 2E and 2F have randomly initialized transforms. Comparing to 1A and other configurations in Setup 2, setups with trainable NWFs exhibit improvements in STOI ranging from 4.44\% to 6.48\%. Across all experiments in Setup 2, optimal performance consistently occurs when calculating the loss using only NWF's output.

Comparing 2C with 2E and 2D with 2F reveals a relatively minor impact of initialization. This implies a potential effective training strategy for $\mathrm{DNN}_1$+NWF is to randomly initialize NWF and jointly train $\mathrm{DNN}_1$ and NWF using a loss at NWF output. Notably, NWF without trainable transforms (MCWF) performs significantly worse than 1A, underscoring the importance of optimizing analysis and synthesis transforms. Finally, objective scores from $\mathrm{DNN}_1$+NWF are similar to those of 1A, suggesting NWF's potential to match a DNN performance while reducing nonlinear distortions.

\subsection{Dual-Stage Setups}
\label{ssec:dual_stage}
In setup 3, we assess the performance of stacking two DNNs, $\mathrm{DNN}_1$ and $\mathrm{DNN}_2$, without a NWF. We train all stages from scratch in 3A and 3B, while in 3C and 3D, we pretrain $\mathrm{DNN}_1$ and subsequently train $\mathrm{DNN}_2$ either independently or jointly with $\mathrm{DNN}_1$. These experiments indicate that the loss in the first stage is not beneficial, and pretraining $\mathrm{DNN}_1$ is crucial for performance improvement. However, the results also reveal that incorporating multiple DNNs offers limited improvements.

%First, we evaluate setup 3 that stacks two DNNs, $\mathrm{DNN}_1$  and $\mathrm{DNN}_2$ without a NWF. In 3A and 3B, we train all stages from scratch, while in 3C and 3D we pretrain $\mathrm{DNN}_1$ and then train $\mathrm{DNN}_2$ with or without $\mathrm{DNN}_1$ together. From 3A and 3B, we conclude that the stage one loss is not useful. The comparison between 3B and 3C exemplifies the importance of pretraining $\mathrm{DNN}_1$, motivating us to establish setup 4 experiments. 3D also has $\mathrm{DNN}_1$ pretrained, but the performance is the worst in setup 3 since $\mathrm{DNN}_1$ does not adapt to $\mathrm{DNN}_2$.
%Experiments conducted in setup 3 shows that stacking multiple DNNs appears to offer limited improvements. %Despite the slight increase in computational resources, 3C exhibits lower performance compared to 1B.

Setups 4 and 5 explore various pretraining methods for stacking $\mathrm{DNN}_1$, NWF, and $\mathrm{DNN}_2$. In setup 4, 4A and 4B utilize MCWF, while 4C and 4D employ NWF. Results reveal superior performance of NWF over MCWF, with the optimal outcome observed when all components are jointly trained in 4D. Similarly, in setup 5, where $\mathrm{DNN}_1$ is pretrained with Wiener filter, MCWF exhibits inferior performance compared to NWF. In 5B to 5D, we introduce random initialization of NWF in pretraining, followed by training NWF with $\mathrm{DNN}_2$ and subsequently training $\mathrm{DNN}_1$ and NWF with $\mathrm{DNN}_2$. While 5B and 5C yield comparable scores, 5D achieves the highest scores, aligning with the observed trend in setup 4.

In light of the insight that joint training leads to better performance, we conducted experiments in Setup 6, involving three distinct loss configurations and no pretraining. A noteworthy observation from 6A to 6C suggests that hitting the "sweet spot" for peak performance involves using the simplest setup—utilizing solely the final output for loss and jointly training all components from scratch.

Our best system (6C) outperforms the baseline (1C) with comparable computational resources, demonstrating improvements in terms of STOI, PESQ, and SI-SDR by $1.82\%$, $0.07$, and $1.14$ dB, respectively, while employing only $56.63\%$ of the parameters. Moreover, in comparison to 6C, model 1E requires $4.22$ and $2.23$ times more parameters and FLOPS to achieve a similar STOI score. Additionally, the best baseline model, FSB-LSTM, is surpassed by employing a larger LLRNN in both stages with a hidden size of $200$. Remarkably, this improved performance is attained with reduced computational requirements, half the algorithmic latency, and a comparable number of parameters.

\section{Conclusion}
\label{sec:conclusion}
We have introduced a novel and resource-efficient framework for sequential neural beamforming in the time-domain, specifically designed for speech enhancement. Within this framework, we have incorporated a novel Neural Wiener Filter (NWF) to enhance low-latency speech processing. We have identified that the most effective training strategy involves simultaneous training of all components, with the final stage's output being used for loss computation. Our best-performing system outperformed robust baseline models across several key metrics, including speech quality, intelligibility, model size, and computational efficiency.

% % Below is an example of how to insert images. Delete the ``\vspace'' line,
% % uncomment the preceding line ``\centerline...'' and replace ``imageX.ps''
% % with a suitable PostScript file name.
% % -------------------------------------------------------------------------
% \begin{figure}[htb]

% \begin{minipage}[b]{1.0\linewidth}
%   \centering
%   \centerline{\includegraphics[width=8.5cm]{image1}}
% %  \vspace{2.0cm}
%   \centerline{(a) Result 1}\medskip
% \end{minipage}
% %
% \begin{minipage}[b]{.48\linewidth}
%   \centering
%   \centerline{\includegraphics[width=4.0cm]{image3}}
% %  \vspace{1.5cm}
%   \centerline{(b) Results 3}\medskip
% \end{minipage}
% \hfill
% \begin{minipage}[b]{0.48\linewidth}
%   \centering
%   \centerline{\includegraphics[width=4.0cm]{image4}}
% %  \vspace{1.5cm}
%   \centerline{(c) Result 4}\medskip
% \end{minipage}
% %
% \caption{Example of placing a figure with experimental results.}
% \label{fig:res}
% %
% \end{figure}

%\vfill
%\pagebreak

% \vfill\pagebreak

\bibliographystyle{IEEEbib}
\bibliography{strings,refs}

\begin{thebibliography}{10}

\bibitem{qian2018deep}
K.~Qian, Y.~Zhang, S.~Chang, X.~Yang, D.~Florencio, and M.~Hasegawa-Johnson,
\newblock ``Deep learning based speech beamforming,''
\newblock in {\em ICASSP}, 2018.

\bibitem{luo2019fasnet}
Y.~Luo, C.~Han, N.~Mesgarani, E.~Ceolini, and S.-C. Liu,
\newblock ``{FaSNet: L}ow-latency adaptive beamforming for multi-microphone
  audio processing,''
\newblock in {\em ASRU}, 2019.

\bibitem{luo2022time}
Y.~Luo,
\newblock ``A time-domain real-valued generalized {W}iener filter for
  multi-channel neural separation systems,''
\newblock {\em IEEE/ACM Transactions on Audio, Speech, and Language
  Processing}, vol. 30, pp. 3008--3019, 2022.

\bibitem{wang2022stft}
Z.-Q. Wang, G.~Wichern, S.~Watanabe, and J.~Le~Roux,
\newblock ``{STFT-domain} neural speech enhancement with very low algorithmic
  latency,''
\newblock {\em IEEE/ACM Transactions on Audio, Speech, and Language
  Processing}, vol. 31, pp. 397--410, 2022.

\bibitem{lu2022towards}
Y.-J. Lu, S.~Cornell, X.~Chang, W.~Zhang, C.~Li, Z.~Ni, Z.-Q. Wang, and
  S.~Watanabe,
\newblock ``Towards low-distortion multi-channel speech enhancement: The
  espnet-se submission to the l3das22 challenge,''
\newblock in {\em ICASSP}, 2022.

\bibitem{kuang2023three}
K.~Kuang, F.~Yang, J.~Li, and J.~Yang,
\newblock ``Three-stage hybrid neural beamformer for multi-channel speech
  enhancement,''
\newblock {\em The Journal of the Acoustical Society of America}, vol. 153, no.
  6, pp. 3378--3378, 2023.

\bibitem{lee2023improved}
C.-H. Lee, C.~Yang, Y.~Shen, and H.~Jin,
\newblock ``Improved mask-based neural beamforming for multichannel speech
  enhancement by snapshot matching masking,''
\newblock in {\em ICASSP}, 2023.

\bibitem{wang2023tf}
Z.-Q. Wang, S.~Cornell, S.~Choi, Y.~Lee, B.-Y. Kim, and S.~Watanabe,
\newblock ``{TF-GridNet: I}ntegrating full- and sub-band modeling for speech
  separation,''
\newblock {\em IEEE/ACM Transactions on Audio, Speech, and Language
  Processing}, vol. 31, pp. 3221--3236, 2023.

\bibitem{wang2018spatial}
Z.-Q. Wang and D.~Wang,
\newblock ``On spatial features for supervised speech separation and its
  application to beamforming and robust {ASR},''
\newblock in {\em ICASSP}, 2018.

\bibitem{ochiai2017unified}
T.~Ochiai, S.~Watanabe, T.~Hori, J.~R Hershey, and X.~Xiao,
\newblock ``Unified architecture for multichannel end-to-end speech recognition
  with neural beamforming,''
\newblock {\em IEEE Journal of Selected Topics in Signal Processing}, vol. 11,
  no. 8, pp. 1274--1288, 2017.

\bibitem{zhang2022end}
W.~Zhang, X.~Chang, C.~Boeddeker, T.~Nakatani, S.~Watanabe, and Y.~Qian,
\newblock ``End-to-end dereverberation, beamforming, and speech recognition in
  a cocktail party,''
\newblock {\em IEEE/ACM Transactions on Audio, Speech, and Language
  Processing}, vol. 30, pp. 3173--3188, 2022.

\bibitem{movsner2022multi}
L.~Mo{\v{s}}ner, O.~Plchot, L.~Burget, and J.~H {\v{C}}ernock{\`y},
\newblock ``Multi-channel speaker verification with conv-tasnet based
  beamformer,''
\newblock in {\em ICASSP}, 2022, pp. 7982--7986.

\bibitem{dowerah2022compensating}
S.~Dowerah, R.~Serizel, D.~Jouvet, M.~Mohammadamini, and D.~Matrouf,
\newblock ``Compensating noise and reverberation in far-field multichannel
  speaker verification,''
\newblock 2022.

\bibitem{wang2020multi}
Zhong-Qiu Wang and DeLiang Wang,
\newblock ``Multi-microphone complex spectral mapping for speech
  dereverberation,''
\newblock in {\em ICASSP}, 2020, pp. 486--490.

\bibitem{wang2023fneural}
Z.-Q. Wang, S.~Cornell, S.~Choi, Y.~Lee, B.-Y. Kim, and S.~Watanabe,
\newblock ``Neural speech enhancement with very low algorithmic latency and
  complexity via integrated full-and sub-band modeling,''
\newblock in {\em ICASSP}, 2023.

\bibitem{doclo2010acoustic}
S.~Doclo, S.~Gannot, M.~Moonen, and A.~Spriet,
\newblock ``Acoustic beamforming for hearing aid applications,''
\newblock {\em Handbook on array processing and sensor networks}, pp. 269--302,
  2010.

\bibitem{warsitz2007blind}
E.~Warsitz and R.~Haeb-Umbach,
\newblock ``Blind acoustic beamforming based on generalized eigenvalue
  decomposition,''
\newblock {\em IEEE Transactions on Audio, Speech, and Language Processing},
  vol. 15, no. 5, pp. 1529--1539, 2007.

\bibitem{mestre2003diagonal}
Xavier Mestre and Miguel~A Lagunas,
\newblock ``On diagonal loading for minimum variance beamformers,''
\newblock in {\em International Symposium on Signal Processing and Information
  Technology}. IEEE, 2003, pp. 459--462.

\bibitem{zhang2021adl}
Zhuohuang Zhang, Yong Xu, Meng Yu, Shi-Xiong Zhang, Lianwu Chen, and Dong Yu,
\newblock ``{ADL-MVDR}: All deep learning {MVDR} beamformer for target speech
  separation,''
\newblock in {\em ICASSP}, 2021.

\bibitem{zhang2020end}
J~Zhang, C.~Zoril{\u{a}}, R.~Doddipatla, and J.~Barker,
\newblock ``On end-to-end multi-channel time domain speech separation in
  reverberant environments,''
\newblock in {\em ICASSP}, 2020.

\bibitem{pandey23b_interspeech}
A.~Pandey, K.~Tan, and B.~Xu,
\newblock ``{A Simple RNN Model for Lightweight, Low-compute and Low-latency
  Multichannel Speech Enhancement in the Time Domain},''
\newblock in {\em INTERSPEECH}, 2023.

\bibitem{patel2023ux}
K.~Patel, A.~Kovalyov, and I.~Panahi,
\newblock ``{UX-Net: F}ilter-and-process-based improved {U-Net} for real-time
  time-domain audio separation,''
\newblock in {\em ICASSP}, 2023.

\bibitem{ba2016layer}
J.~L Ba, J.~R Kiros, and G.~E Hinton,
\newblock ``Layer normalization,''
\newblock {\em arXiv preprint arXiv:1607.06450}, 2016.

\bibitem{he2015delving}
K.~He, X.~Zhang, S.~Ren, and J.~Sun,
\newblock ``Delving deep into rectifiers: Surpassing human-level performance on
  imagenet classification,''
\newblock in {\em ICCV}, 2015.

\bibitem{hochreiter1997long}
S.~Hochreiter and J.~Schmidhuber,
\newblock ``Long short-term memory,''
\newblock {\em Neural Computation}, vol. 9, no. 8, pp. 1735--1780, 1997.

\bibitem{higuchi2018frame}
T.~Higuchi, K.~Kinoshita, N.~Ito, S.~Karita, and T.~Nakatani,
\newblock ``Frame-by-frame closed-form update for mask-based adaptive {MVDR}
  beamforming,''
\newblock in {\em ICASSP}, 2018.

\bibitem{reddy2020interspeech}
C.~KA Reddy, V.~Gopal, R.~Cutler, E.~Beyrami, R.~Cheng, H.~Dubey,
  S.~Matusevych, R.~Aichner, A.~Aazami, S.~Braun, et~al.,
\newblock ``The {INTERSPEECH} 2020 deep noise suppression challenge: Datasets,
  subjective testing framework, and challenge results,''
\newblock {\em INTERSPEECH}, 2020.

\bibitem{pandey2022multichannel}
A.~Pandey, B.~Xu, A.~Kumar, J.~Donley, P.~Calamia, and D.~Wang,
\newblock ``Multichannel speech enhancement without beamforming,''
\newblock in {\em ICASSP}, 2022.

\bibitem{pandey2022tparn}
A.~Pandey, B.~Xu, A.~Kumar, J.~Donley, P.~Calamia, and D.~Wang,
\newblock ``{TPARN}: Triple-path attentive recurrent network for time-domain
  multichannel speech enhancement,''
\newblock in {\em ICASSP}, 2022.

\bibitem{kingma2014adam}
D.~P Kingma and J.~Ba,
\newblock ``Adam: A method for stochastic optimization,''
\newblock {\em arXiv preprint arXiv:1412.6980}, 2014.

\bibitem{pandey2021dense}
A.~Pandey and D.~Wang,
\newblock ``Dense {CNN} with self-attention for time-domain speech
  enhancement,''
\newblock {\em IEEE/ACM Transactions on Audio, Speech, and Language
  Processing}, vol. 29, pp. 1270--1279, 2021.

\end{thebibliography}

\end{document}